\documentclass[epj]{svjour}

\usepackage{graphicx}
\usepackage{fancyhdr}
\def\makeheadbox{{
 }}
\def\mytitle{My title} 
\def\myauthors{My name}  
\def\mytype{My type of session}
\def\mysession{My session}

\def\mytitle{MPP as a mechanism for the suppression of
FCNC and CP--violation in the 2HDM} 
\def\myauthors{R.~Nevzorov}    
\def\mytype{Contributed Talk}    
\def\mysession{Alternatives}

\begin{document}
\pagestyle{empty}
{\onecolumn
\vspace*{-10mm}
\begin{flushright}
{CERN-PH-TH/2007-190}\\
\end{flushright}
\vspace*{25mm}

\begin{center}
{
\LARGE
Multiple point principle as a mechanism for the suppression of FCNC\\[3mm] 
and CP--violation phenomena in the 2HDM}\\[8mm]
{\Large
C.~D.~Froggatt$^a$, R.~Nevzorov$^a$\,${}^{1}$,
H.~B.~Nielsen$^b$\\[3mm]
\itshape{$^a$ Department of Physics and Astronomy,}\\[2mm]
\itshape{Glasgow University, Glasgow, Scotland}\\[2mm]
\itshape{$^b$ The Niels Bohr Institute, Copenhagen, Denmark}
}\\[1mm]
\end{center}
\vspace{16cm} 
\footnoterule{\noindent${}^{1}$
On leave of absence from the Theory Department, ITEP, Moscow, Russia}
}

\newpage
\pagestyle{fancyplain}

\twocolumn
\title{Multiple point principle as a mechanism for the suppression of 
FCNC and CP--violation phenomena in the 2HDM}
\author{C.~D.~Froggatt\inst{1}
\and
R.~Nevzorov\inst{1}
\thanks{\emph{Email:} r.nevzorov@physics.gla.ac.uk}
\and
H.~B.~Nielsen\inst{2}
}                     
%
%
\institute{
Department of Physics and Astronomy, University of Glasgow, Glasgow, G12 8QQ, U.K.
\and
The Niels Bohr Institute, Copenhagen, DK--2100, Denmark}
%
\date{}
\abstract{We argue that multiple point principle (MPP) can be used to ensure CP conservation 
and the absence of flavour changing neutral currents within the two Higgs doublet model (2HDM). 
We also discuss Higgs phenomenology in the MPP inspired 2HDM.
\PACS{
      {12.60.Fr}{Extensions of electroweak Higgs sector}   \and
      {14.80.Bn}{Standard-model Higgs bosons} 
     } 
} 

\maketitle
\section{Introduction}
\label{intro}

In spite of the success of the standard model (SM) in accounting for existing experimental data, 
there are several well--motivated theoretical reasons (like the hierarchy problem, the presence of 
dark matter in the Universe, observed neutrino oscillations etc.) to expect new physics beyond it.
New physics beyond the SM generally introduces new sources of CP violation and gives rise to 
flavour changing neutral current (FCNC) transitions that are forbidden at the tree level in the SM. 
Indeed, the violation of CP invariance and the existence of tree--level flavor--changing neutral currents 
are generic features of $SU(2)_W\times U(1)_Y$ theories with two and more Higgs doublets. Potentially 
large FCNC interactions in these models appear, because the diagonalization of the quark mass matrix 
does not automatically lead to the diagonalization of the two or even more Yukawa coupling matrices which 
describe the interactions of Higgs bosons with fermionic matter. Moreover the Higgs potential of the two--Higgs 
doublet model (2HDM) 
\begin{equation}
\begin{array}{rcl}
V&=& m_1^2 |H_1|^2 + m_2^2 |H_2|^2 -\biggl[m_3^2 H_1^{\dagger}H_2+h.c.\biggr]+\\[0mm]
&&+\displaystyle\frac{\lambda_1}{2} |H_1|^4 + \displaystyle\frac{\lambda_2}{2}|H_2|^4+\lambda_3 |H_1|^2 |H_2|^2+\\[0mm]
&&+\lambda_4|H_1^{\dagger}H_2|^2+\biggl[\displaystyle\frac{\lambda_5}{2}(H_1^{\dagger}H_2)^2 + \lambda_6 |H_1|^2\times\\[0mm]
&&\times (H_1^{\dagger}H_2)+\lambda_7 |H_2|^2(H_1^{\dagger}H_2)+h.c. \biggr]
\end{array}
\label{1}
\end{equation}
contains four couplings $m_3^2$, $\lambda_5$, $\lambda_6$ and $\lambda_7$ that 
can be complex inducing CP--violation.

Recent experimental measurements indicate that FCNC processes are highly suppressed, while 
all CP violation effects are very well described by the phase of CKM matrix. Therefore 
the understanding of the origin of the strong suppression of the FCNC and CP violating processes
caused by new interactions are among the major problems in physics beyond the SM. 
Although one can eliminate the violation of CP invariance and tree--level FCNC transitions in the 2HDM 
by imposing discrete $Z_2$ symmetry, such a symmetry leads to the formation of domain walls in the early 
Universe Ref. \cite{Zeldovich:1974uw} which create unacceptably large anisotropies in the cosmic microwave 
background radiation Ref. \cite{Vilenkin:1984ib}. 

Here, instead of the custodial $Z_2$ symmetry, we use the multiple point principle (MPP) to suppress FCNC and 
CP--violation effects in the 2HDM. The MPP postulates the existence of the maximal number 
of phases with the same energy density which are allowed by a given theory Refs. \cite{mpp1}--\cite{mpp2}. 
The application of the multiple point principle to the SM leads to the remarkable prediction for the top quark (pole) 
and Higgs boson masses Ref. \cite{Froggatt:1995rt}:
\begin{equation}
M_t=173\pm 5\,\mbox{GeV}\, ,\qquad M_H=135\pm 9\, \mbox{GeV}\,,
\label{2}
\end{equation}
which are consistent with current experimental data. In previous papers (see Refs. \cite{sugra1}--\cite{sugra2}) 
the MPP assumption has been adapted to models based on $(N=1)$ local supersymmetry -- supergravity, that allowed 
an explanation for the small deviation of the cosmological constant from zero. Recently we also considered the application 
of the MPP to the SUSY inspired 2HDM of type II Ref. \cite{Froggatt:2006zc}. Here we extend this analysis to the 
general 2HDM and discuss the quasi--fixed point scenario in the MPP inspired 2HDM.

\section{MPP conditions}
\label{sec:2}

The self--consistent implementation of the MPP in the 2HDM can be achieved if,
at some high energy scale $\Lambda$, the following set of degenerate vacua is realized 
(see Refs. \cite{Froggatt:2006zc}--\cite{z2hdm})
\begin{equation}
\langle H_1\rangle=\left(
\begin{array}{c}
0\\ \Phi_1
\end{array}
\right)\,,
\qquad \langle H_2 \rangle=\left(
\begin{array}{c}
0\\ \Phi_2\, e^{i\omega}
\end{array}
\right)\,,
\label{6}
\end{equation}
where $\Phi_1^2+\Phi_2^2=\Lambda^2$ and $\omega$ is an arbitrary parameter.
According to the MPP, vacua at the scale $\Lambda$ must have approximately the same 
energy density as the physical one. This means that all couplings at the MPP scale $\Lambda$ 
have to be adjusted, so that the energy density of the MPP scale vacua vanishes with 
relatively high accuracy. When $\Lambda\gg v$ the mass terms in the 2HDM effective potential 
can be safely ignored, which simplifies our analysis. Then, in compliance with the MPP assumption, 
the $\lambda_i(\Lambda)$ should be adjusted so that an appropriate cancellation takes place
among the quartic terms in the Higgs effective potential (\ref{1}). 

MPP implies that the quartic part of the Higgs potential (\ref{1}) goes to zero for all possible 
values of the phase $\omega$. At the same time for minima to exist, one has also to ensure 
that all partial derivatives of this part of the Higgs potential (\ref{1}) vanish for any choice 
of $\omega$ near the scale $\Lambda$. These two requirements are fulfilled only if
\begin{equation}
\lambda_5(\Lambda)=\lambda_6(\Lambda)=\lambda_7(\Lambda)=0\,,
\label{3}
\end{equation}
\begin{equation}
\beta_{\lambda_5}(\Lambda) = \beta_{\lambda_6}(\Lambda)\Phi_1^2 + \beta_{\lambda_7}(\Lambda)\Phi_2^2 = 0\,, 
\label{4}
\end{equation}
\begin{equation}
\tilde{\lambda}(\Lambda)=\beta_{\tilde{\lambda}}(\Lambda)=0\,,
\label{5}
\end{equation}
where $\tilde{\lambda}(\Phi)=\sqrt{\lambda_1(\Phi)\lambda_2(\Phi)}+\lambda_3(\Phi)+\lambda_4(\Phi)$
while $\beta_{\lambda_i}(\Phi) =\displaystyle\frac{d \lambda_i(\Phi)}{d \ln \Phi}$ is the 
beta function for $\lambda_i(\Phi)$, i.e. we assume here that the $m_i^2$ and $\lambda_i$ depend only on
the overall Higgs norm $\Phi=\sqrt{\Phi_1^2+\Phi_2^2}$.

Eqs.~(\ref{3})--(\ref{5}) represent a complete set of the MPP conditions. These conditions are satisfied 
identically in the minimal SUSY model (MSSM) at any scale lying higher than the masses of the superparticles. 
The MPP conditions (\ref{3})--(\ref{5}) should be supplemented by the vacuum stability requirements:
\begin{equation}
\lambda_1(\Phi)>0,\qquad \lambda_2(\Phi)>0,\qquad \tilde{\lambda}(\Phi)>0,
\label{7}
\end{equation}
which must be satisfied everywhere between the MPP and electroweak scales. Otherwise another minimum of the 
Higgs effective potential arises at some intermediate scale, destabilising the physical and MPP scale vacua.
Taking into account the MPP conditions (\ref{3})--(\ref{5}) and substituting the vacuum configuration (\ref{6}) 
into the quartic part of the 2HDM scalar potential, one finds that near the MPP scale vacua:
\begin{equation}
\Phi_1=\Lambda\cos\gamma,\,\,\,\, \Phi_2=\Lambda\sin\gamma,\,\,\,\,
\tan\gamma=\Biggl(\displaystyle\frac{\lambda_1}{\lambda_2}\Biggr)^{1/4}.
\label{8}
\end{equation}
It is also worth noting that the set of degenerate vacua (\ref{6}) is realised only when
$\lambda_4(\Lambda)$ is negative.

\section{MPP and custodial symmetries}
\label{sec:3}

The MPP conditions constrain the couplings of the Higgs fields to fermions.
The observed mass hierarchy of quarks and charged leptons implies that the Yukawa interactions
in the quark and lepton sectors have a hierarchical structure. Assuming that the Yukawa couplings of the quarks 
and leptons of the third generation are considerably larger than quark and lepton Yukawa couplings of the 
first two generations, the 2HDM Lagrangian describing the interactions of quarks and leptons with the Higgs 
doublets $H_1$ and $H_2$ reduces to
\begin{equation}
\begin{array}{l}
\mathcal{L}'\simeq h_t(H_2\varepsilon 
Q)\bar{t}_R+g_b(H_2^{\dag}Q)\bar{b}_R
+g_{\tau}(H_2^{\dag}L)\bar{\tau}_R
\end{array}
\label{7}
\end{equation}
$$
+g_t(H_1\varepsilon Q)\bar{t}_R+h_b(H_1^{\dag}Q)\bar{b}_R
+h_{\tau}(H_1^{\dag}L)\bar{\tau}_R+h.c.\,,
$$
where $Q$ and $L$ are left--handed doublets of quarks and leptons of the third generation, while $\tau_R$, 
$t_R$ and $b_R$ are right--handed $SU(2)_W$ singlet components of $\tau$--lepton, $t$-- and $b$--quarks.
It is always possible to choose a basis in the field space in which $g_t(\Lambda)=0$. In this basis 
the MPP conditions (\ref{4}) are fulfilled simultaneously only if
\begin{equation}
\begin{array}{cl}
(I)\,& h_b(\Lambda)=h_{\tau}(\Lambda)=0\,;\\[1mm] 
(II)\,& g_b(\Lambda)=g_{\tau}(\Lambda)=0\,;\\[1mm]
(III)\,& h_b(\Lambda)=g_{\tau}(\Lambda)=0\,;\\[1mm] 
(IV)\,& g_b(\Lambda)=h_{\tau}(\Lambda)=0\,.
\end{array}
\label{8}
\end{equation}
The solutions $(I)-(IV)$ correspond to the 2HDM Model I (where only $H_1$ couples to the fermions)
and Model II (where the couplings of $H_1$ and $H_2$ to quarks and leptons are the same as in the MSSM) 
Yukawa couplings and their leptonic variations.

Usually the existence of a large set of degenerate vacua is associated with an enlarged global 
symmetry of the Lagrangian of the considered model. The 2HDM is not an exception. In all models 
$(I-IV)$ the quartic part of the Higgs effective potential (\ref{1}) and $\mathcal{L}'$
are invariant under extra $U(1)$ (Peccei--Quinn) symmetry transformations, which prevent the 
appearance of FCNC transitions at the tree level. 

The generalisation to the three family case requires more accurate consideration. When three generations of 
quarks and leptons have non--negligible couplings to the Higgs doublets, all SM bosons and fermions 
contribute to the Higgs effective potential which is convenient to present in the following form 
\begin{equation}
V_{eff}(H_1, H_2)=\sum_{n=0}^{\infty}V_n(H_1, H_2),
\label{9}
\end{equation}
where $V_0$ corresponds to the tree--level Higgs potential, while $V_n$ represents the
n--loop contribution to $V_{eff}$. In the one--loop approximation we have
\begin{equation}
V_1=\displaystyle\frac{1}{64\pi^2}Str\,|M|^4\biggl[\log\frac{|M|^2}{\mu^2}-C\biggr].
\label{10}
\end{equation}
Here the supertrace operator counts positively (negatively) the number of degrees of freedom 
for the different bosonic (fermionic) fields, $C$ is a diagonal matrix which depends on the
renormalization scheme while $\mu$ is a renormalization scale. Before we restricted
our consideration to the leading log approximation, i.e. we replaced $\log\displaystyle\frac{|M|^2}{\mu^2}$
by $\log\displaystyle\frac{\Phi^2}{\mu^2}$ and summed all leading logs using the renormalisation group 
equations. A more accurate analysis requires us to include all terms that are proportional 
$\Phi^4$ in Eqs.~(\ref{9})--(\ref{10}). 

The independence of the full Higgs effective potential (\ref{9}) on $\omega$ at the scale $\Lambda$ implies 
that the Lagrangian for the Higgs--fermion interactions is invariant under symmetry transformations 
(see Ref. \cite{z2hdm}):
\begin{equation}
\begin{array}{rclcrcl}
H_1 &\to & e^{i\alpha}\,H_1, &\qquad\quad & u'_{R_i}&\to &e^{i\alpha}\,u'_{R_i}, \\
d'_{R_i} &\to & e^{-i\alpha}\,d'_{R_i}, &\qquad\quad & e'_{R_i}&\to &e^{-i\alpha}\,e'_{R_i},\\[3mm]
H_2 &\to & e^{i\beta}\,H_2, &\qquad\quad & u''_{R_i}&\to &e^{i\beta}\,u''_{R_i}, \\
d''_{R_i}&\to & e^{-i\beta}\,d''_{R_i},&\qquad\quad &e''_{R_i}&\to &e^{-i\beta}\,e''_{R_i},
\end{array}
\label{11}
\end{equation}
where $u'_{R_i}$, $d'_{R_i}$, $e'_{R_i}$ are right--handed quarks and leptons which couple to $H_1$ 
while $u''_{R_i}$, $d''_{R_i}$, $e''_{R_i}$ are right--handed fermions that interact with $H_2$.
The renormalisation group (RG) flow of Yukawa couplings does not spoil the invariance of the Lagrangian, 
describing the interactions of quarks and leptons with the Higgs bosons under the custodial 
symmetry transformations (\ref{11}). The two global $U(1)$ symmetries (\ref{11}) guarantee that each 
fermion eigenstate couples to only one Higgs doublet, which in turn ensures the absence of tree--level FCNC transitions. 
Thus MPP provides a reliable mechanism for the suppression of FCNC processes. The MPP solutions based on 
the custodial symmetries (\ref{11}) may be considered as generalisations of the well-known Peccei--Quinn 
symmetric solution of the FCNC problem in the 2HDM. 

Being spontaneously broken at the electroweak scale, the custodial symmetries (\ref{11}) give rise to a 
massless axion which allows us to avoid CP--violation in the 2HDM, entirely eliminating  
the $\theta$--term in QCD. The mixing term $m_3^2(H_1^{\dagger}H_2)$ in the Higgs effective potential (\ref{1}), 
which is not forbidden by the MPP, softly breaks the extra $U(1)$ global symmetry. It spoils the solution 
of the strong CP problem in QCD, but does not create new sources of CP--violation or FCNC transitions. 
Indeed, in the Higgs sector of the general 2HDM only imaginary parts of $m_3^2$, $\lambda_5$, $\lambda_6$
and $\lambda_7$ cause CP--nonconservation. MPP suppresses the Higgs couplings, which are
responsible for the violation of the CP--invariance. At the same time the complex phase of $m_3^2$ can be easily
absorbed by the appropriate redefinition of the Higgs fields. In such a way MPP protects the CP--invariance within
the two Higgs doublet extension of the SM. Tree--level FCNC transitions also do not emerge after
the soft breakdown of the custodial symmetries (\ref{11}), simply because the structure of the interactions of
quarks and leptons with the Higgs doublets remains intact.

There is one important feature that may distinguish the MPP inspired 2HDM from other two Higgs 
doublet models, where softly broken Peccei--Quinn (or $Z_2$) custodial symmetry is postulated.
In the MPP inspired 2HDM the Higgs and Yukawa couplings which violate custodial symmetries can have 
non--zero values, because the multiple point principle does not imply that physical and MPP scale vacua 
should be exactly degenerate. Since we ignore all mass terms in the 2HDM potential (\ref{1}) during the 
derivation of the MPP conditions, one can expect to get the degeneracy of vacua with the accuracy $O(v^2\Lambda^2)$.
This determines the allowed interval of variations of $\tilde{\lambda}(\Lambda)$, $\lambda_{5,\,6,\,7}(\Lambda)$
and custodial symmetry violating Yukawa couplings $g_i$:
\begin{equation}
\begin{array}{c}
|\lambda_5(\Lambda)|\simeq |\lambda_6(\Lambda)|\simeq |\lambda_7(\Lambda)|\le \displaystyle\frac{v^2}{\Lambda^2},\\[2mm]
|g_i(\Lambda)|\le (4\pi)^2 \displaystyle\frac{v^2}{\Lambda^2}.
\end{array}
\label{12}
\end{equation}
Custodial symmetry violating Yukawa and Higgs couplings being set small at the scale $\Lambda$ does not change
much at any scale below $\Lambda$. If $\Lambda$ is quite close to the Planck scale then $\lambda_{5,\,6,\,7}$
and $g_i$ are extremely suppressed at the electroweak scale, so that FCNC and CP--violation effects could not be 
observed in the nearest future. However if $\Lambda\simeq 100\,\mbox{TeV}$ then custodial symmetry violating 
Yukawa couplings may induce non--diagonal flavour transitions, which give rise to new channels of rare 
decays of heavy quarks and leptons that can be detected at future experiments.

\section{Higgs phenomenology}
\label{sec:4}

At moderate values of $\tan\beta=v_2/v_1$, where $v_2$ and $v_1$ are vacuum expectation values of $H_2$ and $H_1$ in the 
physical vacuum, the MPP conditions (\ref{5}) can be rewritten in the following form  
\begin{equation}
\begin{array}{rcl}
\lambda_3(\Lambda) &=&-\sqrt{\lambda_1(\Lambda)\lambda_2(\Lambda)}-\lambda_4(\Lambda)\,,\\[1mm]
\lambda_4^2(\Lambda)&=&\displaystyle\frac{6h_t^4(\Lambda)\lambda_1(\Lambda)}{(\sqrt{\lambda_1(\Lambda)}+\sqrt{\lambda_2(\Lambda)})^2}
-2\lambda_1(\Lambda)\lambda_2(\Lambda)\\[1mm]
&-&\displaystyle\frac{3}{8}\biggl(3g_2^4(\Lambda)+2g_2^2(\Lambda)g_1^2(\Lambda)+g_1^4(\Lambda)\biggr)\,.
\end{array}
\label{13}
\end{equation}
Thus, in the MPP inspired 2HDM, $\lambda_3$ and $\lambda_4$ are not independent parameters. 
As a result the considered 2HDM has less free parameters than the 2HDM of type II and therefore can be regarded as a minimal 
non--supersymmetric two Higgs doublet extension of the SM. 

As follows from Eq.~(\ref{13}), the RG flow of all couplings in the MPP inspired 2HDM is determined by 
$\lambda_1(\Lambda)$, $\lambda_2(\Lambda)$ and $h_t(\Lambda)$. When $h_t(\Lambda)> 1$ the solutions of the RG equations for the top quark 
Yukawa coupling are concentrated in the vicinity of the quasi--fixed point at the electroweak scale. The value
of $\tan\beta$ that corresponds to the quasi--fixed point scenario depends mainly on the MPP scale $\Lambda$.
It varies from $1.1$ to $0.5$ when $\Lambda$ changes from $M_{Pl}$ to $10\,\mbox{TeV}$ \cite{qfp}. At large values of $h_t(\Lambda)$, 
the MPP and vacuum stability conditions constrain $\lambda_i(\Lambda)$ very strongly. 
Our numerical studies show that, for $\Lambda=M_{Pl}$ and 
$\lambda_1(M_{Pl})=\lambda_2(M_{Pl})=\lambda_0$, the ratio $\lambda_0/h_t^2(M_{Pl})$ can vary only within a very narrow 
interval from $0.79$ to $0.87$ if $h_t(\Lambda)> 1.5$. This ensures the convergence of the solutions of the RG equations for 
$\lambda_i(\mu)$ to the quasi--fixed points.

\begin{figure}
\hspace{0cm}{$m_{h_i,\,\chi^{\pm}}$}\\
\includegraphics[width=0.45\textwidth, keepaspectratio=true]{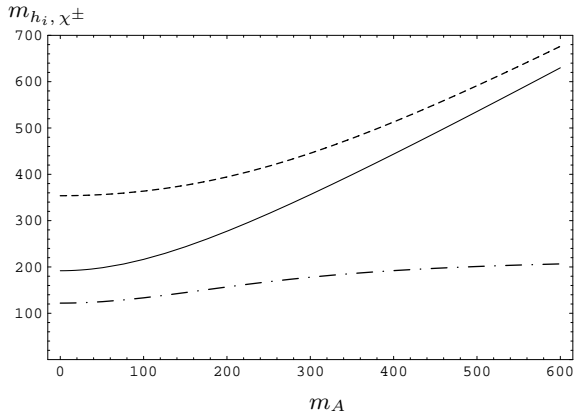}
\hspace*{4cm}{$m_A$}
\caption{Spectrum of Higgs bosons near the quasi--fixed point versus $m_A$ for $\Lambda=10\,\mbox{TeV}$.
The dash--dotted and dashed lines correspond to the CP--even Higgs boson masses while the solid line represents
the mass of the charged Higgs states.}
\label{fig:1}
\end{figure}

\begin{figure}
\hspace{0cm}{$|R_{t\bar{t}h_i}|$}\\
\includegraphics[width=0.45\textwidth, keepaspectratio=true]{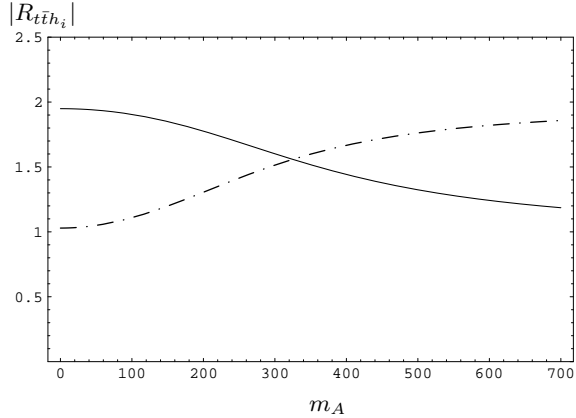}
\hspace*{4cm}{$m_A$}
\caption{Absolute values of the relative couplings $R_{t\bar{t}i}$ of the Higgs scalars to
the top quark in the quasi--fixed point scenario as a function of $m_A$ for $\Lambda=10\,\mbox{TeV}$. 
Solid and dashed--dotted curves correspond to the lightest and heaviest CP--even Higgs states.}
\label{fig:2}
\end{figure}

The Higgs spectrum of the two Higgs doublet extension of the SM contains two charged and three neutral scalar states. 
Because in the MPP inspired 2HDM CP--invariance is preserved, one of the neutral Higgs bosons is always CP--odd while 
the two others are CP--even. The qualitative pattern of the Higgs spectrum depends very strongly on the mass of the 
pseudoscalar Higgs boson $m_A$ (see Fig.~1). When $m_A\gg M_t$ the masses of the charged, CP--odd and heaviest CP--even
Higgs bosons are almost degenerate around $m_A$. In the considered limit the lightest CP--even Higgs boson
mass $m_{h_1}$ attains its maximal value, which is determined by $\lambda_i(M_t)$ and $\tan\beta$. 

For each MPP scale $\Lambda$, one can compute the values of the Higgs self--couplings at the electroweak scale 
and $\tan\beta$ near the quasi--fixed point. Because, at large values of the pseudoscalar mass, $m_{h_1}$ is almost independent 
of $m_A$, the upper bound on the lightest Higgs scalar mass depends predominantly on the scale $\Lambda$. When the MPP scale 
is high such dependence is relatively weak. If $\Lambda> 10^{10}\,\mbox{GeV}$ the mass of the lightest Higgs particle 
in the quasi--fixed point scenario does not exceed $125\,\mbox{GeV}$ \cite{qfp}. 
However at low MPP scales, $\Lambda\simeq 10-100\,\mbox{TeV}$, the theoretical upper limit on $m_{h_1}$ reaches $200-220\,\mbox{GeV}$. 
The lightest Higgs scalar in the considered case is predominantly a SM--like Higgs boson, since its coupling 
to a $Z$ pair is rather close to the SM one. Nevertheless at low MPP scales the quasi--fixed point scenario leads to large 
values of the coupling of the lightest Higgs scalar to the top quark, resulting in the enhanced production 
of this particle at hadron colliders (see Fig.~2). Thus the analysis of production and decay rates of the SM--like Higgs boson at 
the LHC should make possible the distinction between the quasi--fixed point scenario in the MPP inspired 2HDM with low 
scale $\Lambda$, the SM and the MSSM even if extra Higgs states are relatively heavy, i.e. $m_A\simeq 500-700\,\mbox{GeV}$.

\section{Conclusions}
\label{sec:5} 
We have considered the application of the multiple point principle to the non--supersymmetric two-Higgs doublet extension 
of the SM. A complete set of the MPP conditions that leads to the realisation of the MPP assumption has been presented.
We have shown that the existence of a large set of degenerate vacua at some high energy scale $\Lambda$, caused 
by the MPP, results in approximate custodial symmetries which suppress FCNC and CP violating 
phenomena in the 2HDM. 

The spectrum and couplings of the Higgs bosons in the MPP inspired 2HDM have been also studied. When $h_t(\Lambda)>1$ 
the solutions of the RG equations in the considered model converge to the quasi--fixed point, leading to stringent 
restrictions on the lightest Higgs boson mass. In the quasi--fixed point scenario the Higgs couplings to 
the $t$--quark can be significantly larger than in the SM, which allows us to test this scenario at hadron colliders.

\end{document}